# Magnetic and transport properties of Mo substituted La$_{0.67}$Ba$_{0.33}$Mn$_{1-x}$Mo$_x$O$_3$ perovskite system

Darshan C. Kundaliya[*]
*Center for Superconductivity Research, Department of Physics, University of Maryland, College Park, MD-20742, U.S.A.*

Reeta Vij
*Department of Physics, Saurashtra University, Rajkot 360 005, India*

R.G. Kulkarni
*Department of Physics, Shivaji University, Kolhapur-416004, India*

B. Varughese
*Department of Chemistry, University of Maryland, College Park, MD 20742, U.S.A.*

A.K. Nigam and S.K. Malik
*Tata Institute of Fundamental Research, Colaba, Mumbai 400 005, India*

## Abstract

The effect of doping Mo for Mn on the magnetic and transport properties of the colossal magnetoresistance material, La$_{0.67}$Ba$_{0.33}$MnO$_3$, has been studied. Compounds of the series La$_{0.67}$Ba$_{0.33}$Mn$_{1-x}$Mo$_x$O$_3$ (x=0.0 to 0.10) have been prepared and found to crystallize in the orthorhombic structure (space group *Pbnm*). Energy Dispersive X-ray Analysis (EDAX) measurements confirm the stoichiometry of all the samples. Magnetotransport and magnetization measurements reveal that the metal-insulator transition temperature (T$_p$) decreases from 330K for x=0 to 255K for x=0.1. The change in T$_p$ on Mo substitution is relatively much smaller than the corresponding change observed on substitution by other transition elements, such as Ti, Fe, Co, Ni, etc. Further, the ferromagnetic transition temperature (T$_C$) is nearly unchanged by Mo substitution. This is in striking contrast to the large decrease in T$_C$ observed with substitution of above-mentioned 3d elements. These unusual magnetic and transport properties of La$_{0.67}$Ba$_{0.33}$Mn$_{1-x}$Mo$_x$O$_3$ may be either due to the formation of magnetic pair between Mn and Mo or due to strong Mo(4d)-O(2p) overlap, which in turn, may affect the Mn-Mn interaction via the oxygen atoms.

*Corresponding Author: darshan@squid.umd.edu   Phone: +1-301-405-7672

## I. INTRODUCTION

The hole-doped manganese-based perovskite oxides, such as $Ln_{1-x}A_xMnO_3$, (Ln=La, Nd, Pr, etc. and A=Ca, Sr, Ba, Pb, etc.), have been recently the focus of a large number of experimental and theoretical studies. The interest in these materials is because of their extraordinary magnetic, electronic and structural properties including giant negative magnetoresistance [1-3], charge ordering [4-7] and transport properties [8]. Undoped $LaMnO_3$ is an A-type antiferromagnetic insulator [9-11]. By substituting a divalent cation ($A^{2+}$) for La, it can be driven into a metallic and ferromagnetic state. A mixed valence of $Mn^{3+}/Mn^{4+}$ is needed for both metallic behavior and ferromagnetism in these materials. The ferromagnetic interaction between $Mn^{3+}$ and $Mn^{4+}$, caused by the hopping of $e_g$ electrons between two partially filled d orbitals of neighboring $Mn^{3+}$ and $Mn^{4+}$ ions via the orbital overlap $e_g(Mn)$-$2p_\sigma(O)$-$e_g(Mn)$, and a strong on-site Hund's coupling between the $t_{2g}$ core spins and mobile $e_g$ electrons (known as double exchange interaction), plays an important role in determining the magnetic behavior of these compounds[12].

Though there have been numerous studies of the effects of doping at the rare earth site in these manganites, it is only recently that the influence of the substitutions at the Mn site have started attracting some more attention. It has been shown that the trivalent and tetravalent elements substitution for Mn strongly affects the electronic, transport and magnetic behavior of these compounds [13-17]. It has been also observed [17] that, irrespective of the element substituted at the Mn site, the transition temperature ($T_C$) separating the ferromagnetic metallic state and the paramagnetic insulating state decreases dramatically, but the saturated magnetic moment at low temperatures decreases only slightly. In contrast to these observations, studies on mixed valent $Ru^{+4/+5}$ substituted $La_{0.7}A_{0.3}Mn_{0.9}Ru_{0.1}O_3$ (A=Ca, Sr, Ba, and Pb) [18-19] have revealed that the magnetic transition temperature is not much influenced by this substitution. This has been attributed to an unusual pair formation between $Ru^{+4}/Ru^{+5}$ and $Mn^{+4}/Mn^{+3}$. Since Mo can also exist in more than one valence state, it is of interest to study the effect of Mo substitution for Mn on the magnetic and transport properties of such manganites. In this paper, we present the results of such studies on $La_{0.67}Ba_{0.33}Mn_{1-x}Mo_xO_3$ compounds.

## II. EXPERIMENTAL DETAILS

The polycrystalline samples of $La_{0.67}Ba_{0.33}Mn_{1-x}Mo_xO_3$, with x=0.0, 0.025, 0.05, 0.075 and 0.1, were prepared through the solid-state reaction route. The well-mixed stoichiometric proportions of $La_2O_3$, $BaCO_3$, $MnO_2$ and $MoO_3$ (each at least 99.9% pure) were calcinated at 1000°C in air for 24 hrs and reground. The calcinations process was repeated twice. Finally, the calcinated powder was ground, pelletized and sintered at 1200°C for 24 hrs. To check the phase purity of the samples, powder X-ray diffraction measurements were performed using CuKα radiation. The chemical characterization of the samples was performed using Energy Dispersive X-ray Analysis (EDAX) (INCA 200, Oxford). The electrical resistance was measured in the temperature range of 15K to 300K using the four-probe dc method. Magnetoresistance of the samples was also obtained in the temperature range of 5K to 300K in applied magnetic fields upto 9T for x=0.025, 0.075 and 0.10 samples. The magnetization measurements were carried out using a SQUID magnetometer (MPMS, Quantum Design). The X-ray photoelectron spectroscopic (XPS) measurements are done using Kratos Axis 165 spectrometer at a vacuum of $4 \times 10^{-10}$ Torr with monochromatic AlKα radiation. The X-ray power used for the measurements is 150 W. The sample is introduced into the chamber using a conductive carbon tape mounted on a stub. All measurements are done in hybrid mode using both electrostatic and magnetic lenses, with a step size of 0.1 eV and sweep time of 60 s. Survey spectrum is only 3 scans with pass energy of 160 eV. All individual region spectra are recorded in the FAT analyzer mode with pass energy of 20 eV and an average of 10 scans. Charge neutralizer was off during the measurements and binding energy calibration is done with respect to C 1s at 284.6 eV.

## III. RESULTS AND DISCUSSION

The X-ray diffraction (XRD) patterns of all the $La_{0.67}Ba_{0.33}Mn_{1-x}Mo_xO_3$ samples (x=0.0 to 0.1) showed sharp peaks which could be indexed on the basis of an orthorhombically distorted perovskites structure (space group *Pbnm, No. 62)*. No impurity phases were detected with in the x-ray diffraction limits. In order to ascertain the Mo content in these samples, EDAX measurement was carried out. These measurements confirmed the cationic compositions of the samples and, since no secondary phases are seen in the x-rays, it is reasonable to assume that Mo-has substituted for Mn in these samples. A typical EDAX plot (for x=0.025 sample) is shown in Fig. 1(a) and the results of the analysis is largely summarized in Table I.

Rietveld refinements of a typical X-ray diffraction data for x=0.050 sample is shown in Fig. 1(b). The unit cell parameters of the $La_{0.67}Ba_{0.33}Mn_{1-x}Mo_xO_3$ samples, obtained from the Rietveld refinement of the observed XRD patterns, are listed in Table-II. There is a small decrease in the *c*-lattice parameter and an increase in the *a* and the *b* lattice parameters between the parent (x=0.0) and the Mo substituted samples. The unit cell volume also increases with increasing Mo content (Fig. 1(c)). This is expected since the $Mo^{4+/6+}$ ions have slightly larger ionic radii than $Mn^{3+/4+}$ ions. It is worth mentioning that the tolerance factor [defined as $t=\{(R_{Ln}+R_{[O]})/\sqrt{2}(R_{Mn}+R_{[O]})\}$, where R is the ionic radius of the ion concerned] does not change drastically on Mo substitution, and this is important for maintaining the Mn-O-Mn interaction. For instance, the tolerance factor of the parent $La_{0.67}Ba_{0.33}MnO_3$ is about 0.9458, while that for the Mo-substituted sample with x=0.10, is 0.9454.

In order to investigate the magnetic behavior of the $La_{0.67}Ba_{0.33}Mn_{1-x}Mo_xO_3$ samples, magnetization (M) measurements on them were carried out both in the zero-field-cooled (ZFC) state (cooling from 360K) and in the field-cooled (FC) state, in applied fields of 0.005T and 0.5T. Figure 2 shows the plot of ZFC-magnetization ($M_{ZFC}$) versus temperature (T) for all the samples in 0.5T applied field. The inset in Fig.2 shows $M_{ZFC}$ for x=0.025, 0.05, 0.075 and 0.10 samples in low applied field of 0.005T. The Curie temperature ($T_C$) values obtained from the magnetization data of Fig.2 for all the Mo-doped samples are listed in Table-II. It is observed that there is hardly any decrease in $T_C$ for all the $La_{0.67}Ba_{0.33}Mn_{1-x}Mo_xO_3$ samples up to 10 at% Mo-substitution. This is rather unusual compared to the large decrease in $T_C$ observed for other Mn-site substitutions such as Fe, Co, Ni and Ti in such manganites even at the level of 1 at%. The decrease in $T_C$ is more than 130K for 10 at% dopant of such 3d metals and is mainly due to the changes in hole concentration [20,21]. As per Zener's double exchange model, there is neither metallic conduction nor magnetic transition when all the $e_g$ electrons or holes are trapped. The saturation magnetic moment, estimated from the M-H isotherms (Fig. 3), at 5K is approximately 3.25 $\mu_B$/Mn for all the $La_{0.67}Ba_{0.33}Mn_{1-x}Mo_xO_3$ samples. These moments are in good agreement with the free ion values of $Mn^{3+}$ and $Mn^{4+}$ taken in appropriate proportions in the compounds.

Figure 4a shows the resistance ratio ($R_T/R_{295K}$) versus temperature plot for all the $La_{0.67}Ba_{0.33}Mn_{1-x}Mo_xO_3$ samples. Our resistance data on the parent compound are in good agreement with the reported data on $La_{0.67}Ba_{0.33}MnO_3$ [22,23]. For the Mo doped samples, a metal-insulator (M-I) transition takes place at $T_p$ that depends on the level of Mo substitution. The $T_p$ decreases from 330K for x=0 to 235K for x=0.025, to 255K for x=0.05 and thereafter it remains nearly constant around 255K for x=0.05 to 0.1. It may be noted that the $T_p$ values in all the compounds are lower than their $T_C$ values listed in Table-II. Such differences in $T_p$ and $T_C$ are often attributed to the grain boundary effects [22,23]. As we know, grains are nothing but small randomly distributed crystals of varying sizes. When these grains contact each other at a junction, that junction is called a grain boundary. Thus grain boundary is an interface in a single-phase material. It is the region of transition between two crystalline domains (grains) which differ from each other in crystallographic orientation. So on applying a magnetic field, the spin orientations at the grain boundary varies and hence this affects the magnetotransport properties [22,23].

The temperature dependence of the resistance of $La_{0.67}Ba_{0.33}Mn_{1-x}Mo_xO_3$ samples for x=0.025, 0.075 and 0.10, measured in applied fields upto 9T along with the zero-field resistance data. The magnetic field reduces the resistance and shifts $T_p$ towards higher temperatures. The magnetoresistance (MR) ratio, defined as MR=[R(H)-R(0)]/R(0) = $\Delta R/R(0)$, is shown in Figs. 4b, 4c and 4d at selected temperatures for the three samples. In order to reveal the low field MR behavior, data is plotted only upto 1T. It is evident from this figure that the MR increases with decrease in temperature Further, at low temperatures, e.g. at 50K, the MR value increases with increasing field and is found to be 27% in 2T field and 44% in 9T field for x=0.025 sample; 27% (2T) and 44.5% (9T) for x=0.075 sample; and 29% (2T) and 46%(9T) for x=0.10 sample, respectively. Jia *et al.* [24] have discussed the role of grain boundaries in the manganites and according to them, the finite MR obtained for polycrystalline manganite materials for $T \ll T_C$, i.e., at low temperatures, most likely results from grain boundary scattering. We have also observed appreciable low temperature MR at 50K for $La_{0.67}Ba_{0.33}Mn_{1-x}Mo_xO_3$ compounds, which decreases with increasing temperatures, $T \ll T_C$, (Fig. 4), and is believed to be due to grain boundary effects. This is also confirmed by us taking low field MR measurements. Similar MR behaviour has been also observed by Ju *et al.* [22] for the polycrystalline $La_{0.67}Ba_{0.33}MnO_3$ and $La_{0.8}Ca_{0.2}MnO_3$ samples and attributed to the presence of grain boundary effects. Kobayashi *et al.* [25] have explained such MR behavior in terms of spin-polarized intergrain tunneling arising out of grain boundaries.

Further, in order to ascertain the valence state of Mo, XPS measurements were performed and the spectrum for Mo 3d is shown in Fig. 5. Data processing is done using Vision processing software. After substraction of a linear background, spectra are fitted using 60% Gaussian/40% Lorentzian peaks, takin the minimum number of peaks consistent with the best fit the important parameters used for this fitting are peak position, full width at half maximum, intensity and the Gaussian fraction which determines the fraction of the Gaussian component in the fitted peak shape. Interestingly, Mo 3d appears as two doublets, one at 231.2 eV and another at 232.5 eV. Binding energy of 231.2 eV is very close to the literature reported value of +4 ions [27] where as 232.5 eV corresponds to +6 ions [28,29]. These suggest that Mo like Ru also exists in mixed valence states and probably forming a magnetic pair with Mn.

From the above results, it is clear that the Mo substitution at the Mn site in $La_{0.67}Ba_{0.33}MnO_3$ gives entirely different result compared to those obtained by substituting other transition metals at the same site. The present results on Mo substituted samples are similar to those obtained by Ru substitution at the Mn site in which a magnetic pair formation between Ru and Mn moments has been proposed. A similar magnetic pair formation between $Mn^{+3}:Mo^{+6} \leftrightarrow Mn^{+4}:Mo^{+4}$ moments could also take place in the present system. Further, since the 4d electrons are more itinerant relative to the 3$d$ electrons, it is also possible that the 4d substitution (e.g. Mo and Ru) at the 3$d$-Mn site, causes more overlap of Mo/Ru (4d) electrons with the O(2p) electrons, which in turn, modifies the Mn-O-Mn interaction. The Mo (IV) ion either forms a low spin state ($^3T_{1g}:t^4_{2g}e^0$) or a high spin state ($^5A_{1g}:t^3_{2g}e^1$) with effective magnetic moments of 2.83 and 4.90$\mu_B$, respectively. If Mo (IV) would have present in a high spin state, the $T_C$ should have increased for a possible Mn-O-Mo interaction. The very little drop observed in $T_C$ for Mo substitution up to 10 at% doping suggests that Mo (IV) exists in low spin state. Further, to provide charge neutrality, Mo (like Ru) assumes a mixed valence state in these compounds and confirmed by the XPS measurements. From the study of ordered perovskite $Sr_2FeMoO_6$ [25, 26], Mo was assigned to the +5 valence state (with spin of ½) rather than +6 valence state. In view of these, it will be worth exploring the valence and magnetic states of Mo and Ru ions in such compounds by various techniques like neutron scattering.

**References:**

[1] S. Jin, T.H. Tiefel, M. McCormack, R.A. Fastnacht, R. Ramesh, and L.H. Chen, Science **264,** 413 (1994).

[2] R. von Helmolt, J. Wecker, B. Holzapfel, L. Schultz, and K. Samwer, Phys. Rev. Lett. **71,** 2331 (1993).

[3] R. Mahesh, R. Mahendiran, A. K. Raychaudhuri, and C. N. R. Rao, J. Solid State Chem. **114**, 297 (1995).

[4] E. O. Wollan and W. C. Koehler, Phys. Rev. **100**, 545 (1955).

[5] P. Schiffer, A. P. Ramirez, W. Bao, and S.-W. Cheong, Phys. Rev. Lett. **75**, 3336 (1995).

[6] Y. Tomioka, A. Asamitsu, Y. Moritomo, H. Kuwahara, and Y.Tokura, Phys. Rev. Lett. **74**, 5108 (1995).

[7] K. Liu, X. W. Wu, K. H. Ahn, T. Sulchek, C. L. Chien, and J. Q. Xiao, Phys. Rev. B **54**, 3007 (1996).

[8] H. Kuwahara, Y. Tomioka, A. Asamitsu, Y. Moritomo, and Y. Tokura, Science **270**, 961 (1995).

[9] I. Solovyev, N. Hamada, and K. Terakura, Phys. Rev. Lett. **76**, 4825 (1996).

[10] W. E. Pickett and D. J. Singh, Phys. Rev. B **53**, 1146 (1996).

[11] G. H. Jonker and J. H. van Santen, Physica (Amsterdam) **16**, 337 (1950); J. H. van Santen and G. H. Jonker, *ibid.* **16**, 599 (1955).

[12] J. Töpfer and J. B. Goodenough, Chem. Mater. **9**, 1467 (1997).

[13] C. Martin, A. Maignan, and B. Raveau, J. Mater. Chem. **6**, 1245 (1996).

[14] K. H. Ahn, X. W. Wu, K. Liu, and C. L. Chien, J. Appl. Phys. **81**, 5505 (1997).

[15]  J. Blasco, J. Garcia, J. M. de Teresa, M. R. Ibarra, J. Perez, P. A. Algarabel, C. Marquina, and C. Ritter, Phys. Rev. B **55**, 8905 (1997).

[16]  C. Osthover, P. Grunberg, and R. R. Arons, J. Magn. Magn. Mater. **177-181**, 854 (1998).

[17]  A. Maignan, C. Martin, and B. Raveau, Z. Phys. B: Condens. Matter **102**, 19 (1997).

[18]  S.S. Manoharan, H.L. Ju and K.M. Krishan, J. Appl. Phys. **83**, 7183 (1998).

[19]  R.K. Sahu and S.S. Manoharan, Appl. Phys. Lett. **77**, 2382 (2000).

[20]  J. Gutierrez, J. M. Barandiaran, M. Insauati, L.L. A. Pena, J.J. Blanco and T. Rujo, J. Appl. Phys. **83**, 7171 (1998).

[21]  B. Raveau, A. Maignan and C. Martin, J. Solid State Chem. **130**, 162 (1997).

[22]  H.L. Ju and H. Sohn, Solid State Comm. **102,** 463 (1997).

[23]  H.L. Ju, J. Gopalakrishnan, J.L. Peng, Q. Li, G.C. Xiong, T. Venketesan and R.L. Greene, Phy. Rev. B **51**, 6143 (1995).

[24]  Y.X. Jia, Li Lu, K. Khazeni, V.H. Crespi, A. Zettl, Marvin L. Cohen, Phy. Rev. B **52**, 9147 (1995).

[25]  K.-I. Kobayashi, T. Kimura, H. Sawada, K. Terakura and Y. Tokura, Nature (London) **395**, 677 (1998).

[26]  J. Lindén, T. Yamamoto, M. Karppinen, H. Yamauchi and T. Pietari, Appl. Phys. Lett. **76**, (2000) 2925.

[27]  O.G. Samuel and L.J. Matienzo, Inorganic Chemistry **14**, (1975) 1014.

[28]  D.D. Sarma and C.N.R. Rao, J. Electron Spectrosc. Relat. Phenom. **20**, (1980) 25.

[29] C.D. Wagner, W.M. Riggs, L.E. Davis, J.F. Moulder, G.E. Mullenberg, Handbook of X-Ray photoelectron Spectroscopy, Perkin-Elmer Corporation, Physical Electronics Division, Eden prairie, MN 1979.

**Figure Captions**

Figure 1    Typical (a) EDAX plot for $La_{0.67}Ba_{0.33}Mn_{1-x}Mo_xO_3$ sample with x=0.025. (b) Rietveld refinement of X-ray diffraction data for $La_{0.67}Ba_{0.33}Mn_{1-x}Mo_xO_3$ with x=0.050. Open circles are observed x-ray diffraction data and the line is a theoretical fit to the observed x-ray data. Vertical bars are the Bragg reflections for the space group *Pbnm* and below that is a difference pattern of observed data and the theoretical fit. (c) Lattice volume versus Mo content plot for $La_{0.67}Ba_{0.33}Mn_{1-x}Mo_xO_3$ compounds

Figure 2    Plot of magnetic moment (M), measured in a field of 0.5T, versus temperature (T) for the Mo substituted $La_{0.67}Ba_{0.33}Mn_{1-x}Mo_xO_3$ samples Inset shows the temperature variation of ZFC and FC magnetization measured in 0.005T field.

Figure 3    Plot of magnetic moment (M) versus applied field (H) at 5K for $La_{0.67}Ba_{0.33}Mn_{1-x}Mo_xO_3$ samples with x=0.025, 0.05, 0.075 and 0.10.

Figure 4    (a) Variation of resistance ratio ($R_T/R_{295K}$) versus temperature for $La_{0.67}Ba_{0.33}Mn_{1-x}Mo_xO_3$ samples with x=0.0 to 0.10. MR% vs applied field at selected temperatures for $La_{0.67}Ba_{0.33}Mn_{1-x}Mo_xO_3$ samples with (b) x=0.025, (c) x=0.075 and (d) x=0.10.

Figure 5    X-ray photoelectron spectroscopy (XPS) measurement of $La_{0.67}Ba_{0.33}Mn_{0.95}Mo_{0.05}O_3$ shows mixed valent state of Molybdenum.

Table I: Results of EDAX analysis of $La_{0.67}Ba_{0.33}Mn_{1-x}Mo_xO_3$ samples

| Sample 'x' | Element | EDAX | | Nominal Composition |
|---|---|---|---|---|
| | | Wt. % | Stoichiometric composition calculated | |
| 0.025 | O | 20.44 | --- | $3\pm\delta$ |
| | Mn | 21.41 | 0.974 | 0.9750 |
| | Mo | 1.21 | 0.026 | 0.0250 |
| | Ba | 18.70 | 0.340 | 0.33 |
| | La | 38.24 | 0.687 | 0.67 |
| 0.050 | O | 20.45 | --- | $3\pm\delta$ |
| | Mn | 20.85 | 0.948 | 0.950 |
| | Mo | 2.16 | 0.056 | 0.050 |
| | Ba | 18.59 | 0.337 | 0.33 |
| | La | 37.95 | 0.679 | 0.67 |
| 0.075 | O | 20.45 | --- | $3\pm\delta$ |
| | Mn | 20.33 | 0.924 | 0.925 |
| | Mo | 2.81 | 0.073 | 0.075 |
| | Ba | 18.55 | 0.336 | 0.33 |
| | La | 37.86 | 0.680 | 0.67 |
| 0.1 | O | 20.49 | --- | $3\pm\delta$ |
| | Mn | 20.00 | 0.910 | 0.9 |
| | Mo | 4.07 | 0.106 | 0.1 |
| | Ba | 18.40 | 0.335 | 0.33 |
| | La | 37.04 | 0.667 | 0.67 |

Table II: Structural and magnetic properties of Mo-substituted bulk $La_{0.67}Ba_{0.33}Mn_{1-x}Mo_xO_3$ samples. Here x is Mo content, a, b, c are the lattice parameters, t is the tolerance factor [defined as $t=\{(R_{Ln}+R_{[O]})/\sqrt{2}(R_{Mn}+R_{[O]})\}$] and MR% is the magnetoresistance ratio defined as MR%=$\{[R(H)-R(0)]/R(0)\}*100$, $T_C$ is the Curie temperature and $T_p$ is the metal-insulator transition temperature.

| x in Sample | Lattice Parameters[1] (Å) | | | Tolerance Factor[2] (t) | Curie Temperature $T_C$ (K) | M-I Transition[3] $T_p$ (K) | MR% at 50K | |
|---|---|---|---|---|---|---|---|---|
| | a | b | c | | | | 2T | 9T |
| 0.0 | 5.5013 | 5.5013 | 7.6881 | 0.9458 | 340 | 330 | - | - |
| 0.025 | 5.5198 | 5.5029 | 7.6806 | 0.9457 | 340 | 235 | 27 | 44 |
| 0.050 | 5.5718 | 5.5178 | 7.6610 | 0.9456 | 340 | 255 | - | - |
| 0.075 | 5.6026 | 5.5320 | 7.6274 | 0.9455 | 340 | 258 | 27 | 44.5 |
| 0.100 | 5.6244 | 5.5594 | 7.6123 | 0.9454 | 340 | 255 | 29 | 46 |

1) Accuracy ± 0.002 Å
2) Accurate within ± 0.003 Å
3) Accuracy ± 3K

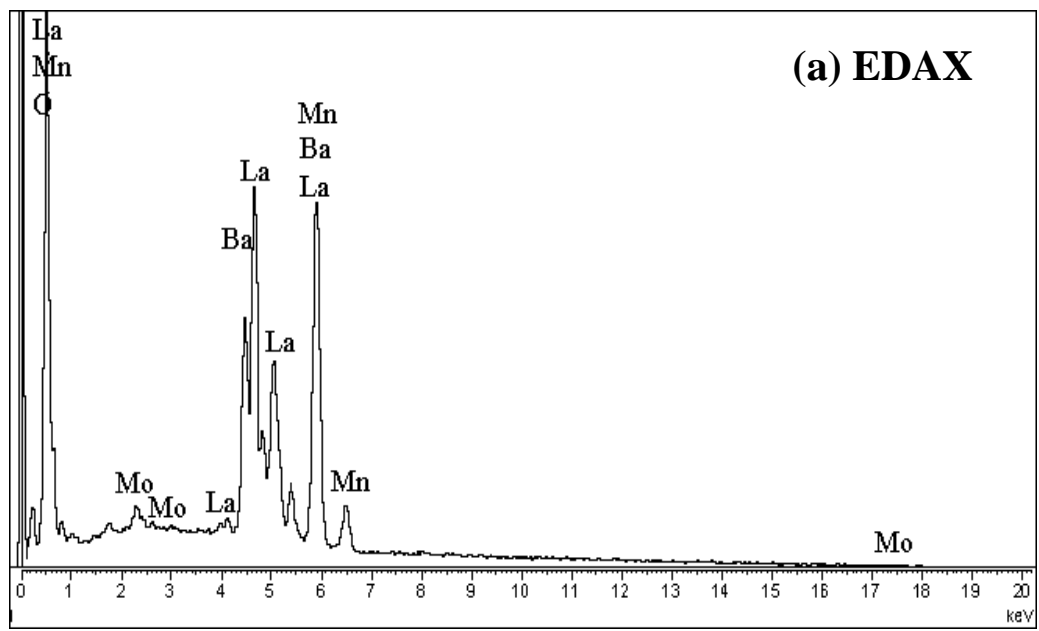

Figure 1 (a) Kundaliya *et al.*

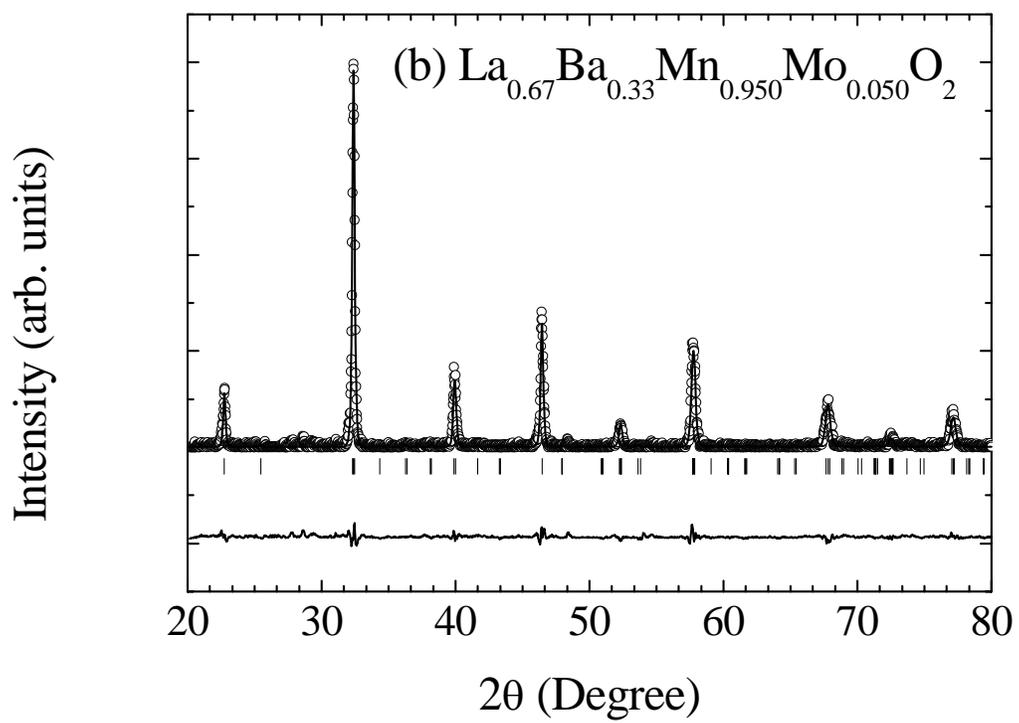

Figure 1 (b) Kundaliya *et al.*

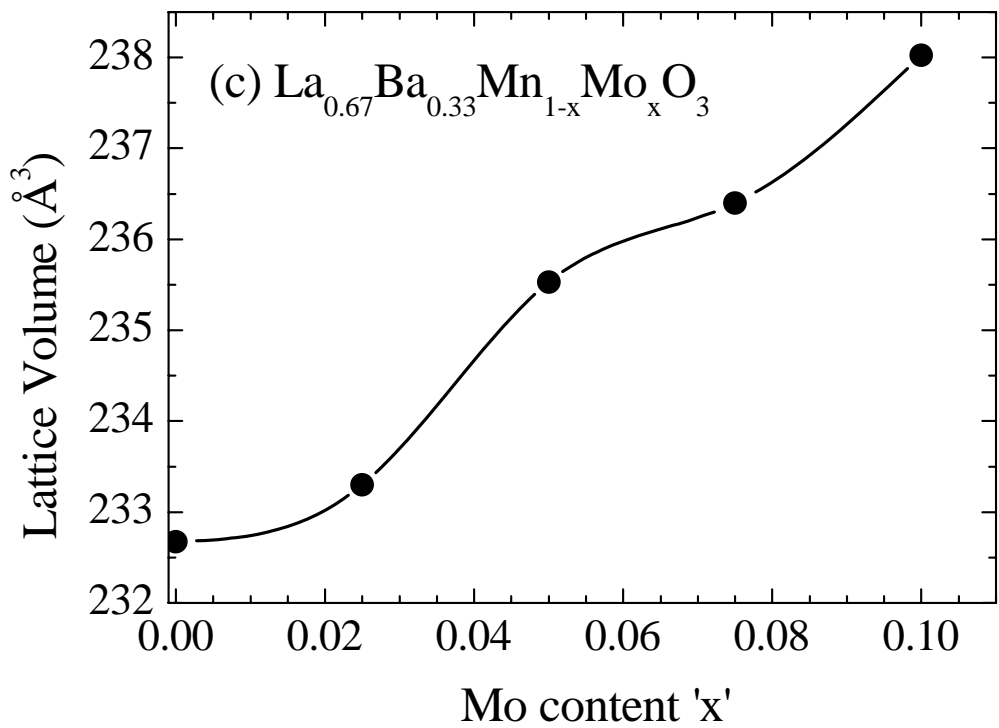

Figure 1 (c) Kundaliya *et al.*

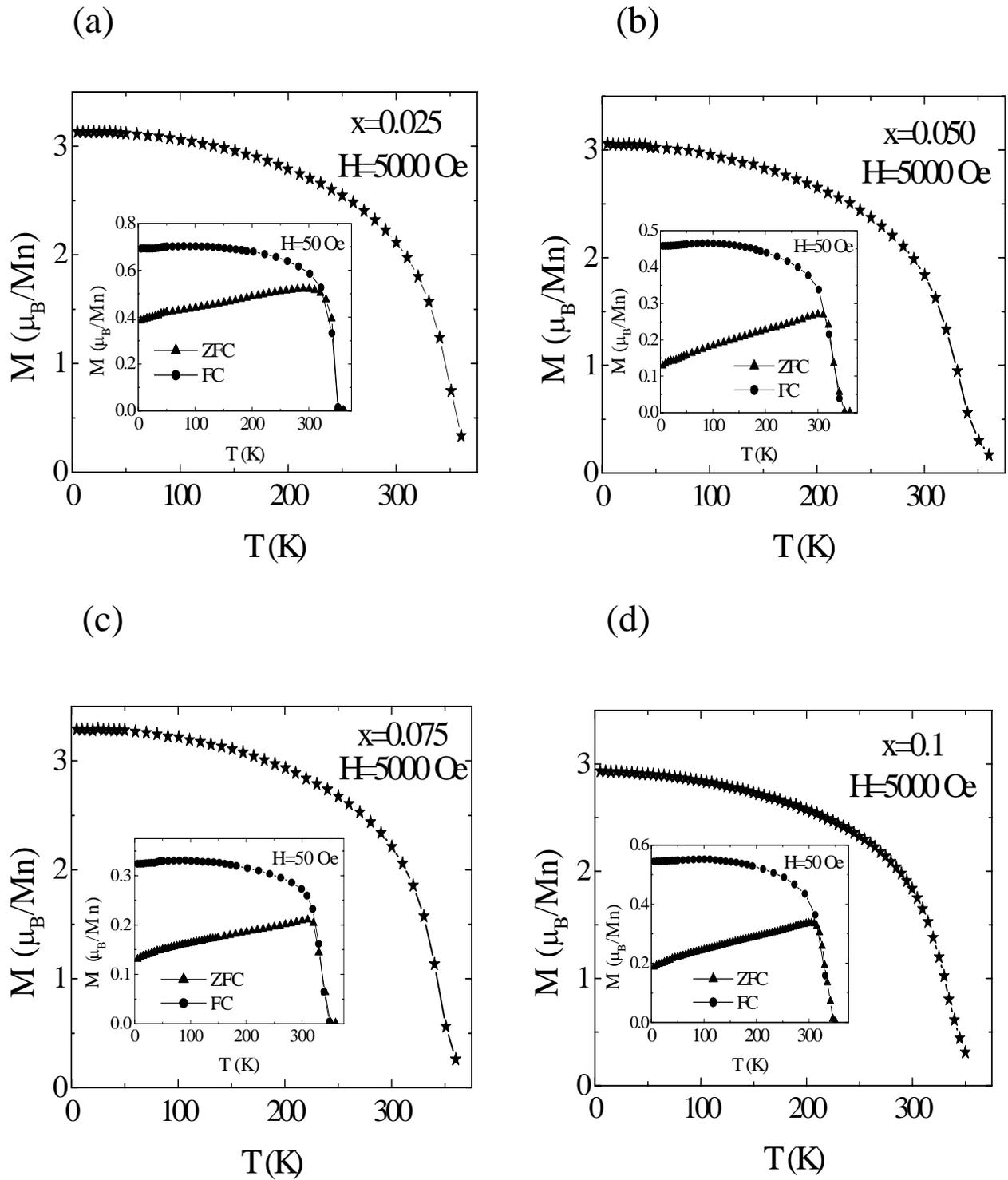

Figure 2 Kundaliya *et al.*

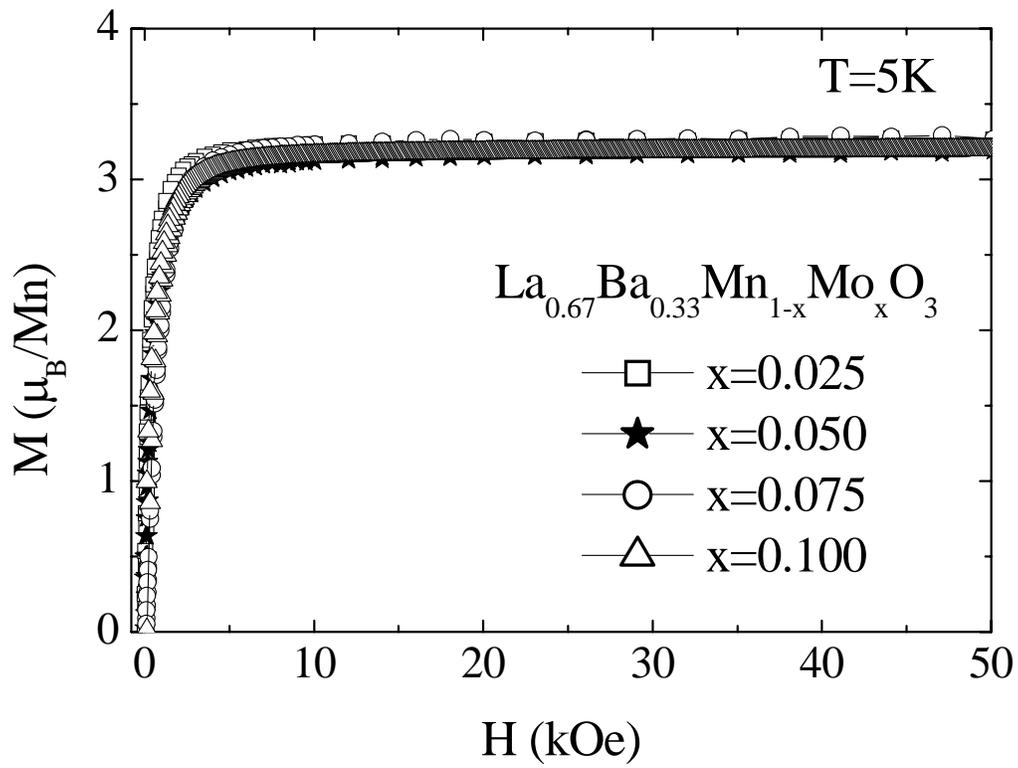

Figure 3 Kundaliya *et al*.

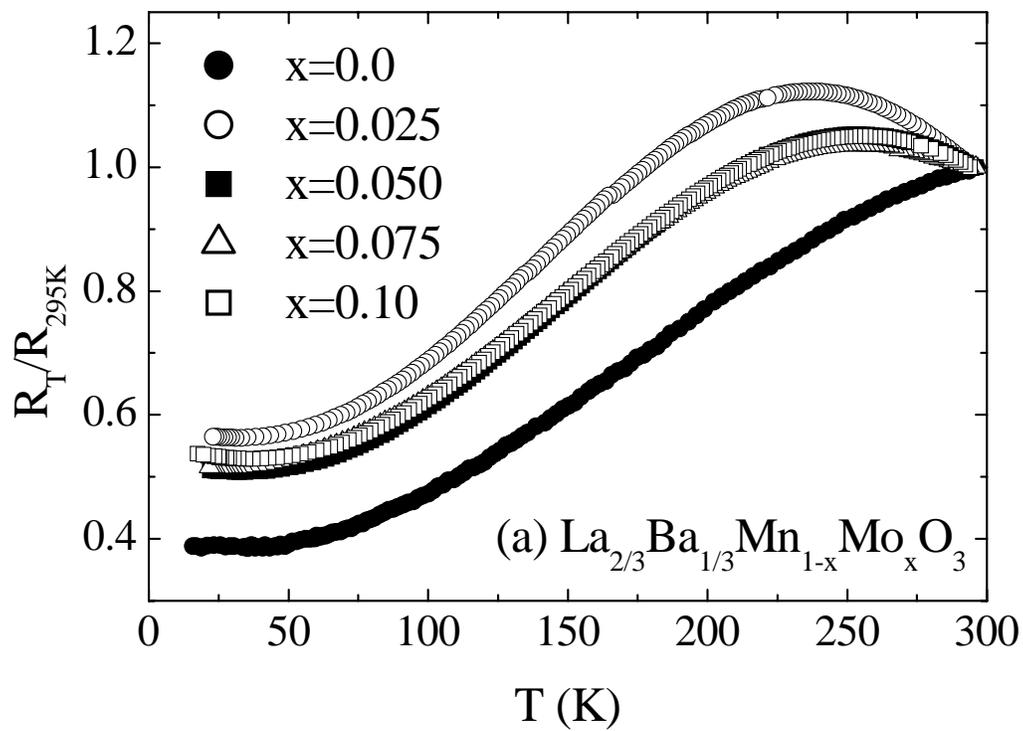

(a) $La_{2/3}Ba_{1/3}Mn_{1-x}Mo_xO_3$

Figure 4a Kundaliya *et al.*

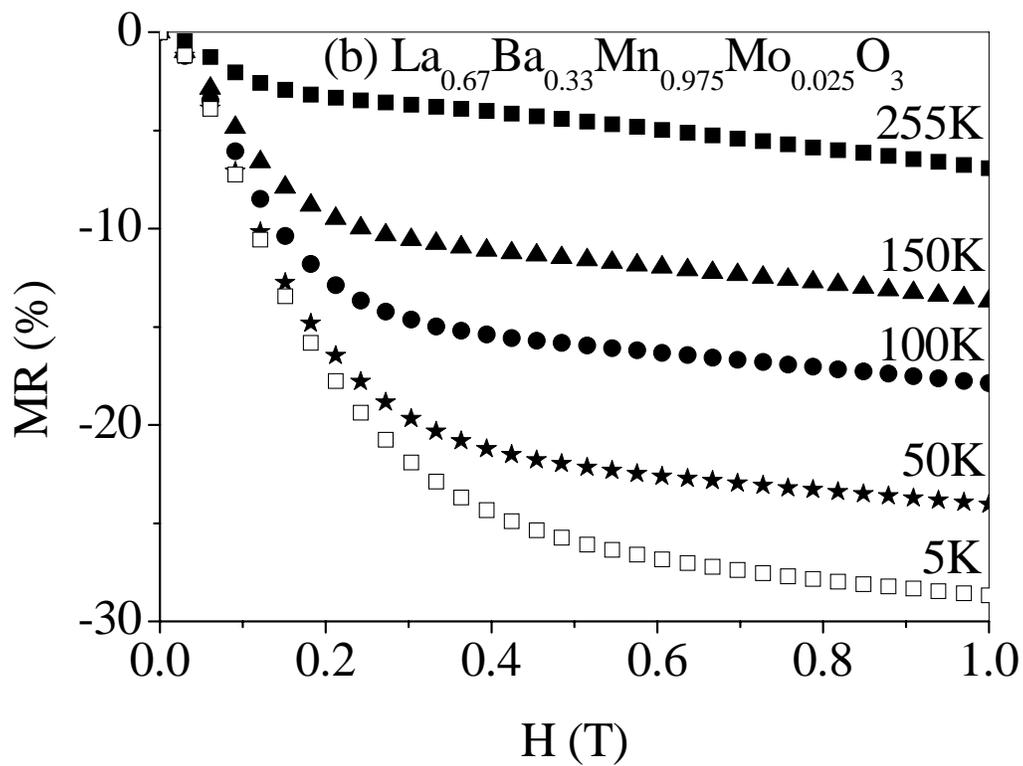

Figure 4b Kundaliya *et al.*

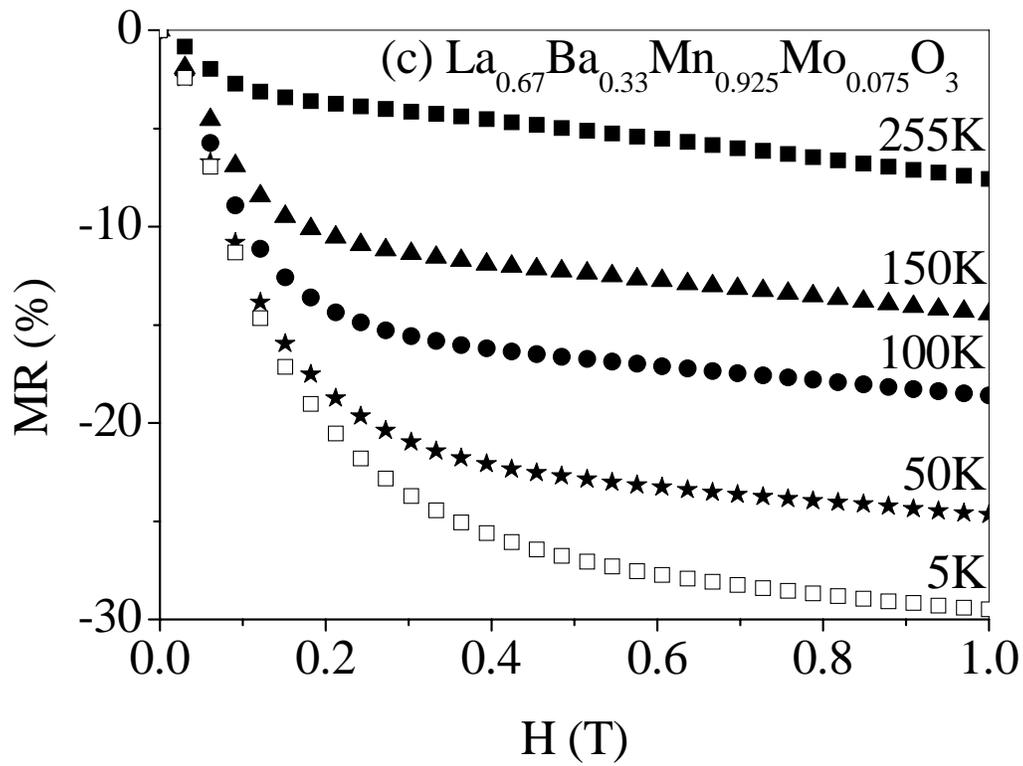

Figure 4c Kundaliya *et al.*

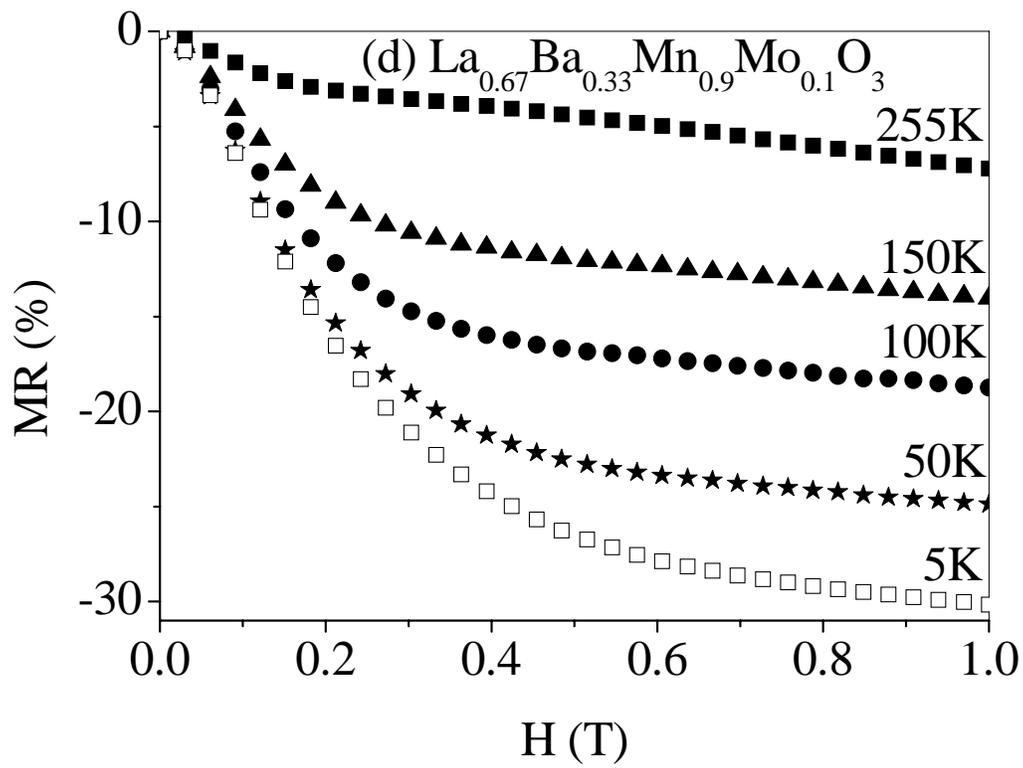

Figure 4d Kundaliya *et al.*

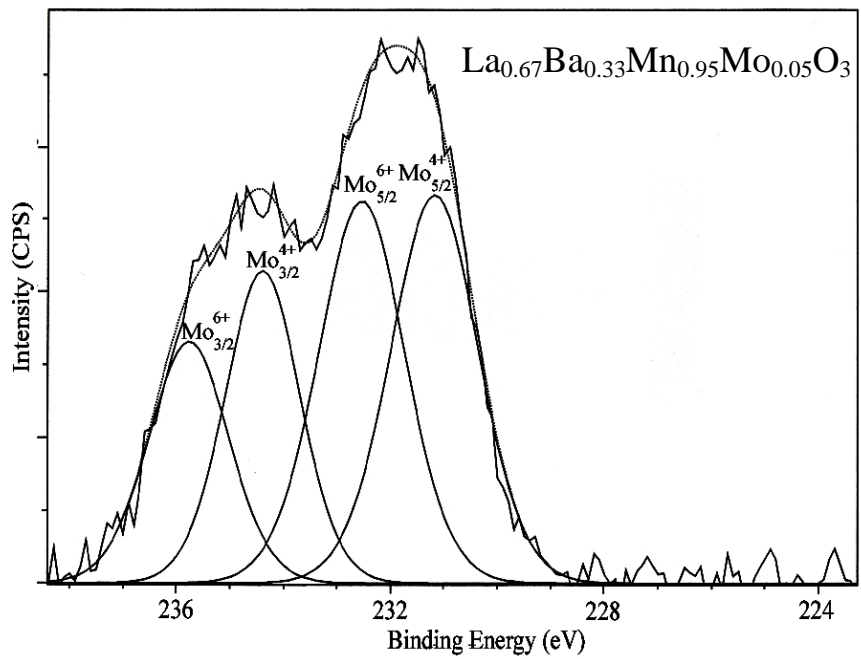

Figure 5 Kundaliya *et al.*